\documentstyle[aps,twocolumn,prb,epsf]{revtex} 

\begin{document}  \preprint{\today} \draft
\title{Single Hole Green's Functions in Insulating
Copper Oxides at Nonzero Temperature}
 
\author{J. van den Brink\cite{G} and O. P. Sushkov\cite{Budker}}

\address{School of Physics, The University of New
South Wales, Sydney 2052, Australia}
\maketitle
 
\begin{abstract}

We consider the single hole dynamics in a modified $t-J$ model
at finite temperature. 
The modified model includes a next nearest ($t^{\prime}$) and next-next 
nearest ($t^{\prime \prime}$) hopping. 
The model has been considered before in the zero temperature limit to
explain angle resolved photo-emission measurements.
We extend this consideration to the case of
finite temperature where long-range anti-ferromagnetic order is destroyed,
using the self-consistent Born approximation.
The Dyson equation which relates the single hole Green's functions
for a fixed pseudo-spin and for fixed spin is derived.
The Green's function with fixed pseudo-spin is infrared stable but 
the Green's function with fixed spin is close to an infrared divergency.
We demonstrate how to renormalize this Green's function in order to assure 
numerical convergence.
At non-zero temperature the quasi-particle peaks are found 
to shift down in energy and to be broadened.
\end{abstract}


\section{Introduction}
Recent angle resolved photo-emission (ARPES) measurements by Wells 
{\it et al}~\cite{8Wells} and by Pothuizen {\it et al}~\cite{8Pothuizen}
on the insulating Copper Oxide Sr$_2$CuO$_2$Cl$_2$ provide an 
unique experimental probe of the properties of a single hole
in an anti-ferromagnetic background.
Theoretically this problem was analyzed in terms of a $t-t^{\prime}-J$ model
using exact diagonalization techniques for small clusters~\cite{8Naz} and 
the self-consistent Born approximation (SCBA)~\cite{8Bala}.
From recent evaluation of the hopping integrals it was concluded that
the next-next nearest neighbor hopping matrix element $t^{\prime \prime}$ is 
significant and almost as large as the diagonal hopping matrix
element $t^{\prime}$, so that $t^{\prime \prime}$ should also 
be incorporated in a model Hamiltonian~\cite{8And}.
This was done in a recent paper~\cite{8G1}, where also the leading
corrections to the SCBA were evaluated.

The ARPES experiments~\cite{8Wells,8Pothuizen} are carried out at a
temperature of 300-350K, which is above the Neel temperature of this
compound. Theoretical treatments up to now, however, restrict themselves
to the zero temperature limit, assuming long range anti-ferromagnetic order,
and spectra are artificially broadened in order to compare with experiment.
It is therefore important to extend the SCBA calculations to finite
temperature, where long range anti-ferromagnetic order is lacking,
although at room temperature the magnetic correlation length for this
compound is still about 60 lattice spacings and no drastic deviation
of the ARPES spectrum at room temperature from the one at zero temperature
is expected.
Another motivation for this work is that a SCBA technique that
can cope with the absence of long range magnetic order may be
extended in the future to describe the spin liquid state of the doped 
copper oxides.

We consider a  two-dimensional 
$t-t^{\prime}-t^{\prime \prime}-J$ model at finite temperature.
We apply the modified spin-wave theory suggested by Takahashi for 2D Heisenberg
model at nonzero temperature~\cite{8Taka} to deal with a state 
without long range anti-ferromagnetic order. 
The Hamiltonian for $t-t^{\prime}-t^{\prime \prime}-J$ model is
\begin{eqnarray}
\label{H}
H&=&-t\sum_{\langle ij \rangle \sigma} c_{i\sigma}^{\dag}c_{j\sigma}
-t^{\prime}\sum_{\langle ij_1 \rangle \sigma}
c_{i\sigma}^{\dag}c_{j_1\sigma}
-t^{\prime \prime}\sum_{\langle ij_2 \rangle \sigma} 
c_{i\sigma}^{\dag}c_{j_2\sigma} \nonumber \\
&+&J \sum_{\langle ij \rangle \sigma} {\bf S}_i{\bf S}_j,
\end{eqnarray}
where $c_{i \sigma}^{\dag}$ is the creation operator of an electron with
spin $\sigma$ $(\sigma =\uparrow, \downarrow)$ at site $i$
on the two-dimensional square lattice, $\langle ij \rangle$ represents
nearest neighbor sites, $\langle ij_1 \rangle$ next nearest neighbor sites
(diagonal), and $\langle ij_2 \rangle$ represents next-next
nearest sites. The spin operator is ${\bf S}_i={1\over 2}
c_{i \alpha}^{\dag} {\bf \sigma}_{\alpha \beta} c_{i \beta}$.
The exchange derived from two magnon Raman scattering
is $J=125meV$~\cite{8Tok,8Grev}. Following the most recent 
calculation of the hopping matrix elements performed by 
Andersen {\it et al}~\cite{8And} we take: $t=386meV$, $t^{\prime}=-105meV$,
$t^{\prime \prime}=86meV$.  We set $J=1$, so that in these units
\begin{equation}
\label{ts}
t=3.1, \ \ \ t^{\prime}=-0.8, \ \ \ t^{\prime \prime}=0.7
\end{equation}

We first calculate the hole
Green's function with fixed pseudo-spin at finite temperature,
introduced by a constraint on the sub-lattice magnetization, and 
evaluate the contribution to the self energy due to the virtual 
absorption of spin-waves by the hole.
Then the hole Green's function with fixed spin, that corresponds
to Green's function that is actually measured in ARPES, is calculated.
This Green's function turns out to be close an infrared divergency and we
show that this instability can be avoided by a proper renormalization,
assuring that the results numerically converge even when a rather
limited number of grid-points is used.
We find that at non-zero temperature the quasi-particle peaks 
broaden and shift to lower energy. The shift is independent of momentum and is
due to the larger effective hole bandwidth as at finite temperature
the spin order is frustrated. 

\section{Hole Green's function with fixed pseudo-spin G$_{\rm d}$}

At half filling (one electron per site)
the model under consideration is equivalent to a Heisenberg
model. We are interested in the situation when one electron is
removed from the system, when a single hole is produced.
The dynamics of a single hole in an anti-ferromagnetic background
can be described by SCBA~\cite{8Schmitt,8Kane}. This 
approximation works very well due to the absence of a single loop 
corrections to the hole-spin-wave vertex~\cite{8Mart,8Liu,8Susf}. 
Now we have to modify SCBA for finite temperature.
The main complication is that at finite temperature there is no
long range anti-ferromagnetic order.
Nevertheless, following Takahashi~\cite{8Taka} we introduce
artificially two sub-lattices: up and down.
The bare hole operator $d_i$ is defined so that $d_{i}^{\dag}
\propto c_{i \uparrow}$ on the $\uparrow$ sub-lattice and 
$\propto c_{i \downarrow}$ 
on the $\downarrow$ sub-lattice. In momentum representation
\begin{eqnarray}
\label{d}
d_{{\bf k}\downarrow}^{\dag} =\sqrt{2\over {N (1/2+m)}}\sum_{i \in \uparrow}
c_{i \uparrow}e^{i{\bf k}{\bf r}_i} \nonumber \\
d_{{\bf k}\uparrow}^{\dag} =\sqrt{2\over {N (1/2+m)}}\sum_{j \in \downarrow}
c_{j \downarrow}e^{i{\bf k}{\bf r}_j},
\end{eqnarray}
where $N$ is number of sites and $m= \langle S_{iz}\rangle =0$ is
the average magnetization.  The brackets $\langle \rangle$ denote
both quantum and statistical averaging. The quasi-momentum
${\bf k}$ is restricted to be inside the magnetic
Brillouin zone: $\gamma_{\bf k}= {1\over 2}(\cos k_x + \cos k_y) \ge 0$.
In this notations it looks like $d_{{\bf k} \sigma}$ has spin
$\sigma =\pm 1/2$, but actually rotation invariance is violated
and $\sigma$ is a pseudo-spin which just labels the two sub-lattices. 
Nevertheless the pseudo-spin gives the correct value of the spin $z$-projection:
$S_z=\sigma =\pm 1/2$. The coefficients in (\ref{d}) provide the correct
normalization:
\begin{equation}
\label{norm}
\langle d_{{\bf k}\downarrow} d_{{\bf k}\downarrow}^{\dag}\rangle
={4\over {N}}\sum_{i \in \uparrow}
\langle c_{i \uparrow}^{\dag}c_{i \uparrow}\rangle
=2 \left({1\over 2}+\langle S_{iz}\rangle \right)=1.
\end{equation}
The retarded hole Green's function is defined as
\begin{equation}
\label{Gd}
G_d(\epsilon,{\bf k})=-i\int_0^{\infty} \langle d_{{\bf k}\sigma}(\tau)
d_{{\bf k}\sigma}^{\dag}(0) \rangle e^{i \epsilon \tau} d\tau
\end{equation}
The $t^{\prime}$, $t^{\prime \prime}$  terms
in the Hamiltonian (\ref{H}) correspond to the 
hole hopping inside one sub-lattice. This gives the bare hole dispersion
\begin{eqnarray}
\label{e0}
\epsilon_{0{\bf k}}&=&4t^{\prime}\cos k_x \cos k_y+
2t^{\prime \prime}(\cos 2k_x + \cos 2k_y) \nonumber \\
&\to& \beta_{01}\gamma^2_{\bf k}+\beta_{02}(\gamma^-_{\bf k})^2,
\end{eqnarray}
where $\gamma^-_{\bf k}={1\over 2}(\cos k_x - \cos k_y)$,
$\beta_{01}=4(2t^{\prime \prime}+t^{\prime})$, and
$\beta_{02}=4(2t^{\prime \prime}-t^{\prime})$.
In equation (\ref{e0}) we took into account that the sign of a hole
dispersion is opposite to that for an electron (the maximum of the electron
band corresponds to the minimum of the hole band), and omitted the constant
term. 
The bare hole Green's function is
\begin{equation}
\label{Gd0}
G_{0d}(\epsilon, {\bf k})={1\over{\epsilon - \epsilon_{0{\bf k}}+i0}}.
\end{equation}

For spin excitations we use the modified spin-wave theory~\cite{8Taka}
(see also review paper~\cite{8Manousakis}).
In order to treat the Heisenberg term in the Hamiltonian (\ref{H}) within
spin-wave theory, it is convenient to use the Dyson-Maleev 
transformation~\cite{8DM} for a localized spin $S=1/2$,
\begin{eqnarray}
\label{dm}
&&S_l^-=a_l^{\dag}, \ \ \ S_l^+=(2S-a_l^{\dag}a_l)a_l,\nonumber\\
&&S_l^z=S-a_l^{\dag}a_l, \ \ \ for\ \ l \in up \ \ \ sublattice;\\
&&S_m^-=b_m, \ \ \ S_m^+=b_m^{\dag}(2S-b_m^{\dag}b_m),\nonumber\\
&&S_m^z=-S+b_m^{\dag}b_m, \ \ \ for\ \ m \in down \ \ \ sublattice,
\nonumber
\end{eqnarray}
and the Fourier representation for $a_l$ and $b_m$:
\begin{eqnarray}
\label{Fu}
a_l&=&\sqrt{2\over{N}}\sum_{\bf q}e^{i{\bf q}{\bf r}_l}a_{\bf q}\\
a_m&=&\sqrt{2\over{N}}\sum_{\bf q}e^{i{\bf q}{\bf r}_m}b_{\bf q}.\nonumber
\end{eqnarray}
The summation over ${\bf q}$, here and everywhere below, is restricted inside 
the magnetic Brillouin zone. 
There are essentially two ways to find an effective Hamiltonian quadratic 
in the operators $a$ and $b$. The first way is just to drop
the quartic terms as is done in linear spin-wave theory (LSWT).
The second way is to treat the quartic terms at mean-field level
$a^{\dag}a b^{\dag}b \to
\langle a^{\dag}a\rangle b^{\dag}b +
\langle a^{\dag}b^{\dag}\rangle  ab + ...$, corresponding to mean field
spin-wave theory. As both approximations give very close
results we choose use LSWT because it is simpler in the present context. 

\subsection{Finite Temperature Correction for G$_d$}

So far we followed the zero temperature derivation for the SCBA.
We take the approach of calculating the finite temperature corrections
to the hole Green's function G$_d$ with a diagrammatic, perturbative,
method. This framework can be used if the number of spin-waves per site
is small, i.e. if $T/J$ is not too large. This is a reasonable 
pre-requisition as in the experiments $T/J$ is of the order of 1/4.
We will show later on that at these temperatures the number of spin-waves
is actually small, justifying our approach.
The advantage of this approach over performing the calculations
at  imaginary, Matsubara, frequencies is that in our framework the absence
of long-range anti-ferromagnetic order is specifically built in. 
Also problems with the numerical analytic continuation on the real frequency
axis are avoided. As is shown later on, an almost-divergency occurs that demands
large lattices in order to get stable results. By performing the
calculation on the real axis from the beginning, the Green's function
can be renormalized, lifting the almost-divergency.

At non-zero temperatures the long range anti-ferromagnetic order is 
destroyed. Very many long-wave-length spin waves are excited, 
and one must take into account their non-linear interaction. 
An approximate way to do it is to impose an additional condition that 
the sub-lattice magnetization is zero~\cite{8Taka}.
\begin{equation} 
\label{constraint}
\langle S^z_{l \in \uparrow} - S^z_{m \in \downarrow} \rangle =
 \langle 
 \frac{1}{2} - a^{\dag}_{\bf l} a_{\bf l}  +
 \frac{1}{2} - b^{\dag}_{\bf m} b_{\bf m} \rangle
 = 0.
\end{equation}
This constraint gives an effective cutoff of unphysical 
states in Dyson-Maleev transformation.
The constraint (\ref{constraint}) is introduced into the Hamiltonian
via a Lagrange multiplier ${1\over 8}\nu^2$. 
Now we must diagonalize
\begin{eqnarray} 
\label{Hnu}
 H_{eff}&=& H_{LSWT} - {1\over 8}\nu^2 (S^z_{l \in \uparrow} - 
S^z_{m \in \downarrow}) \\
&\to & 2\sum_{\bf q}\left(A(a^{\dag}_{\bf q}a_{\bf q}+
b^{\dag}_{\bf q}b_{\bf q})
+\gamma_{\bf q}(a_{\bf q}b_{-\bf q}+a^{\dag}_{\bf q}b^{\dag}_{-\bf q})\right),
\nonumber
\end{eqnarray}
where $A=1+\nu^2/8$.
This can be done by the Bogoliubov transformation
\begin{eqnarray}
\label{Bog}
a_{\bf q}&=&u_{\bf q}\alpha_{\bf q}+v_{\bf q}\beta^{\dag}_{-\bf q},\\
b_{-\bf q}&=&v_{\bf q}\alpha^{\dag}_{\bf q}+u_{\bf q}\beta_{-\bf q},\nonumber
\end{eqnarray}
and we find the effective spectrum and Bogoliubov parameters
\begin{eqnarray}
\label{sw}
&&\omega_{\nu \bf q}=2\sqrt{A^2-\gamma_{\bf q}^2},\nonumber \\
&&u_{\bf q}=\sqrt{{A\over{\omega_{\nu \bf q}}}+{1\over 2}},\\
&&v_{\bf q}=-sign(\gamma_{\bf q})
\sqrt{{A\over{\omega_{\nu \bf q}}}-{1\over 2}}.\nonumber
\end{eqnarray}
These equation show that at non-zero temperature the spin-wave spectrum has 
a gap \ $\nu\sqrt{1+{{\nu^2}\over{16}}} \approx \nu$.
This elucidates the meaning of the constraint and the Lagrange multiplier.
Taking into account that in thermal equilibrium
\begin{equation}
\label{nk}
n_{\bf q}\equiv \langle \alpha^{\dag}_{\bf q}\alpha_{\bf q}\rangle
=\langle \beta^{\dag}_{\bf q}\beta_{\bf q} \rangle =
{1\over{\exp(\omega_{\nu \bf q}/T)-1}}
\end{equation}
we obtain from (\ref{constraint}) the equation for $\nu$
\begin{equation}
\label{nu}
0=1-{2\over N}\sum_{\bf q}
{{A}\over{\omega_{\nu \bf q}}}(1+2n_{\bf q}).
\end{equation}
This equation gives an exponentially small $\nu$, and hence an exponentially
large magnetic correlation length $\xi_M \propto 1/\nu$.

Hopping to a nearest neighbor in the Hamiltonian (\ref{H}) gives an
interaction of the hole with spin-waves. 
\begin{eqnarray}
\label{hsw}
H_{h,sw}&=&\sum_{\bf k,q}g_{\bf k,q} (
d_{{\bf k+q}\downarrow}^{\dag}d_{{\bf k}\uparrow}\alpha_{\bf q} \nonumber \\
&+& d_{{\bf k+q}\uparrow}^{\dag}d_{{\bf k}\downarrow}\beta_{\bf q}+
H.c.).
\end{eqnarray}
In diagrams we will denote the vertex $g_{\bf k,q}$ by a dot, see figure~\ref{fig:Fig3}d.
This vertex is given by
\begin{eqnarray}
\label{g}
g_{\bf k,q} &\equiv& \langle \alpha_{\bf q}d_{{\bf k}\uparrow}|H_t|
d_{{\bf k+q}\downarrow}^{\dag} \rangle\nonumber\\
&=&4t\sqrt{2\over N}(\gamma_{\bf k}u_{\bf q}+
\gamma_{\bf k+q}v_{\bf q}).
\end{eqnarray}
In this calculation the usual mean field
factorization approximation
\begin{eqnarray}
\langle \alpha_{\bf q}c_{j \downarrow}^{\dag}c_{j \uparrow}
c_{i \uparrow}^{\dag}c_{i \uparrow}\rangle \approx
\langle \alpha_{\bf q}c_{j \downarrow}^{\dag}c_{j \uparrow}\rangle
\langle c_{i \uparrow}^{\dag}c_{i \uparrow}\rangle=
1/2 \cdot \langle \alpha_{\bf q}S_j^-\rangle\nonumber
\end{eqnarray}
has been used.
For simplicity we have omitted in the calculation (\ref{g}) the standard 
Bose statistics factor $\sqrt{1+n_{\bf q}}$, but they are certainly
taken into account in the calculation of diagrams.
We stress that the vertex (\ref{g}) has the same form as in the case
of zero temperature, except for the pseudo-gap in the Bogoliubov
parameters (\ref{sw}). We remind of the fact that at zero temperature
$g_{{\bf k,q}=0}=0$ because of the Goldstone theorem.
In the present case due to the pseudo-gap $g_{{\bf k,q}=0}\ne 0$.
This is a reflection of the fact that the long range anti-ferromagnetic 
order is destroyed. 
However, as the pseudo-gap is small, its presence will not give rise to large
effects in the spectra.

\begin{figure}
      \epsfxsize=80mm
      \centerline{\epsffile{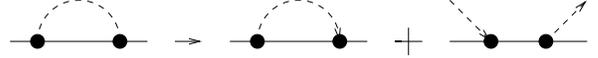}}
\caption{Self energy at finite temperature. The solid line represents 
the hole Green's function $G_d$, and the dashed line represents the spin-wave.
The first diagram corresponds to virtual emission of the spin wave, and 
the second diagram corresponds to virtual absorption.}
\label{fig:Fig1}
\end{figure}

Similar to the zero temperature case the spin structure of the 
interaction (\ref{hsw}) forbids single loop corrections  to the
hole-spin-wave vertex, so that the self energy is of the form
\begin{eqnarray}
\label{SE}
\Sigma(\epsilon, {\bf k})&=&\sum_{\bf q}
[(1+n_{\bf q})g^2_{\bf k-q,q} G_d(\epsilon -\omega_{\bf q},{\bf k-q})  \nonumber\\
&+&  n_{\bf q}g^2_{\bf k,q} G_d(\epsilon +\omega_{\bf q},{\bf k+q}) ]),
\end{eqnarray}
where first term arises from the virtual emission of the spin wave and
second term arises from the virtual absorption, see figure~\ref{fig:Fig1}. The self-consistent 
solution of this equation together with the standard relation
\begin{equation}
\label{Dy}
G_d(\epsilon, {\bf k})={1\over{\epsilon - \epsilon_{0{\bf k}}
-\Sigma(\epsilon, {\bf k}) + i 0}},
\end{equation}
gives the retarded Green's function. Due to the definition of the operators 
(\ref{d}) the Green's function (\ref{Gd}) is invariant under translation  
with the inverse vector of the magnetic sub-lattice ${\bf Q}=(\pm \pi,\pm \pi)$
\begin{equation}
\label{Bl}
G_d(\epsilon,{\bf k+Q})=G_d(\epsilon,{\bf k})
\end{equation}
in spite of the absence of the long range anti-ferromagnetic order.

\subsection{Results for G$_d$}

\begin{figure}
      \epsfxsize=80mm
      \centerline{\epsffile{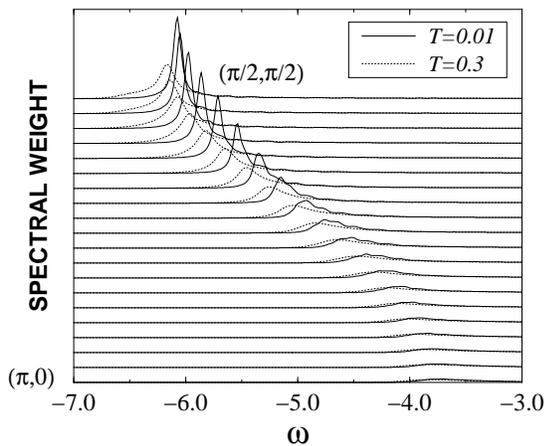}}
\caption{Plots of $-{1\over{\pi}} \ Im \ G_d(\omega, {\bf k})$
as a function of $\omega$ for a cut through the Brillouin zone from 
${\bf k}=(\pi/2, \pi/2)$ to ${\bf k}=(\pi, 0)$ for T=0.01 and T=0.3.}
\label{fig:cut}
\end{figure}

The numerical solution of equation (\ref{Dy}) is straightforward.
To avoid poles we replace $i0 \to i \Gamma/2 =i \ 0.05$. 
The energy scale consists of 300 points with variable density
(concentrated near sharp structures of $G_d$).
The number of points in the magnetic Brillouin zone is $10^4$
which is equivalent to a $140 \times 140$ lattice.
Actually the results are almost independent of the grid as soon as it is
larger than $20 \times 20$.
In figure~\ref{fig:cut} $-{1\over{\pi}} \ Im \ G_d(\omega, {\bf k})$
as a function of $\omega$ for a cut through the Brillouin zone from 
${\bf k}=(\pi/2, \pi/2)$ to ${\bf k}=(\pi, 0)$ is shown for two different
temperatures.

The number of spin-waves per site for the highest temperature, $T=0.4$,
we considered, is $\approx 0.05$, so much smaller than unity, justifying
the perturbational approach.
We recall that we use the set of parameters (\ref{ts}) based on 
Ref.~\cite{8And}. The same set has been used in Ref.~\cite{8G1} for
a zero temperature calculation. Quasiparticle energies and residues 
obtained here are quite similar to that at zero 
temperature~\cite{8G1}. We see from figure~\ref{fig:cut} that temperature shifts
the quasi-peaks positions to lower energy and broadens them. 
The explanation for this is that there are two contributions to the self-energy
at non-zero temperatures. One term originates from virtual spin-wave emission processes
and the other from virtual spin-wave absorption. The former contribution
is also present at T=0 and is enhanced at finite temperatures by a factor $1+n_{\bf q}$,
i.e. at finite temperature the interaction between the spin-waves and the hole is
effectively increased due to this process. 
The matrix element for emission of an extra spin-wave is
larger if the number of spin-waves already present is larger, because of the
bosonic nature of the spin-waves. The part of the self-energy that is due to the
spin-wave emission is multiplied by a factor larger than one at finite temperature, causing
a shift of the quasi-particle peaks to lower energy that is nearly uniform in the Brillouin zone 
and a shift of the incoherent part of the Greens' function to higher energies.
This has a simple physical reason. At non-zero temperature the hole
propagates more easily because the anti-ferromagnetic order is frustrated. This
causes a uniform shift of all quasiparticle poles to lower energy and
does not effect their dispersion, as the quasiparticle dispersion is
determined predominantly by the magnetic interaction. So the effect of
non-zero temperature is qualitatively different from the effects
of doping, where it is found that a reconstruction of the quasiparticle dispersion takes
place, attributed to the frustrated magnetic order in the doped system~\cite{8Ed}.
The contribution to the self-energy at finite temperature because of the spin-wave
absorption is mainly responsible for the broadening of the quasi-particle peaks.

\section{Hole Green's function with fixed spin G$_c$}

The operators $d_{{\bf k}\uparrow}$, $d_{{\bf k}\downarrow}$ 
discussed in the previous section are
defined at different sub-lattices. The operators on two sub-lattices are
useful as mathematical constructions but when a photon removes
an electron from the system, it does not differentiate between the 
sub-lattices, and moreover, at nonzero temperature there are no sub-lattices 
at all.
Therefor we have to define the particle operator that is relevant
for photo-emission independent of the sub-lattice.
This particle operator is:
\begin{equation}
\label{c}
c_{{\bf k}\sigma} =\sqrt{2\over N}\sum_i
c_{i \sigma}e^{i{\bf k}{\bf r}_i}.
\end{equation}
The normalization is chosen in such a way that 
\begin{equation}
\langle 0|c^{\dag}_{{\bf k}\uparrow} c_{{\bf k}\uparrow}|0\rangle=
{2 \over N} \langle 0|\sum_i c^{\dag}_{i\uparrow}c_{i\uparrow} 
|0\rangle=1. \nonumber
\end{equation}
The corresponding retarded Green's function is
\begin{equation}
\label{Gc}
G_c(\epsilon,{\bf k})=-i\int_0^{\infty} \langle c_{{\bf k}\sigma}^{\dag}
(\tau)c_{{\bf k}\sigma}(0) \rangle e^{i \epsilon \tau} d\tau.
\end{equation}
This is the Green's function measured in ARPES.

Now we have to find the relation between $G_c(\epsilon,{\bf k})$
and $G_d(\epsilon,{\bf k})$.
The operator $c_{{\bf k}\sigma}$ acting on half filled ground-state
produces a single hole. We denote the corresponding amplitude by 
$a_{\bf k}$ and denote it in figure~\ref{fig:Fig3}a as a cross. The thick line corresponds 
to the Green's function $G_c$ (\ref{Gc}) and the thin line corresponds to 
the $G_d$ (\ref{Gd}). The amplitude $a_{\bf k}$ equals
\begin{eqnarray}
\label{ak}
a_{\bf k}&=&\langle d_{{\bf k}\uparrow} c_{{\bf k}\downarrow} \rangle 
=\sqrt{{1\over{2}}}.
\end{eqnarray}

The operator $c_{{\bf k}\sigma}$ acting on a state of the system can also 
produce a hole + spin-wave. This amplitude is shown in figure~\ref{fig:Fig3}b
as a circled cross with the dashed line being a spin-wave. We denote this
amplitude by $b_{\bf k,q}$  
\begin{eqnarray}
\label{bk}
b_{\bf k-q,q}&=&\langle \beta_{\bf q} 
d_{{\bf k-q}\downarrow} c_{{\bf k}\downarrow} \rangle\nonumber\\
&=& {{2\sqrt{2}}\over{N}}\langle \beta_{\bf q}
\left( \sum_{i \in \uparrow} S_i^+ e^{i{\bf qr_i}}\right)\rangle \\
&=& {{2\sqrt{2}}\over{N}}\langle \beta_{\bf q}
\left( \sum_{i \in \uparrow}
(1-a_i^{\dag}a_i)a_i e^{i{\bf qr_i}}\right)\rangle
\approx\sqrt{2\over{N}} v_{\bf q} \nonumber.
\end{eqnarray}
There is some ambiguity in the last step of this derivation. If we neglect
the quartic term $\beta_{\bf q}(1-a_i^{\dag}a_i)a_i \to \beta_{\bf q}a_i$ we get
a value of $b_{\bf k,q}$ by a factor $\sqrt{2}$ larger than that given
by eq. (\ref{bk}). If we treat the quartic term on
a mean field level then $a_i^{\dag}a_i \to 1/2$ and we get a value of 
$b_{\bf k,q}$, a factor $\sqrt{2}$ smaller than in eq.(\ref{bk}).
The correct value is somewhere in between. We choose the vertex
$b_{\bf k,q}$ to be the same as in the zero temperature case~\cite{8G1}.
We will see that this provides the correct sum rule for the Green's function
$G_c$, and this is a justification of our choice.
We stress that (\ref{bk}) is a bare vertex. It corresponds to the
instantaneous creation of a hole + spin-wave, but not the creation of a hole
with a subsequent decay into a hole + spin-wave. 

\subsection{Finite Temperature Correction for G$_c$}

\begin{figure}
      \epsfxsize=80mm
      \centerline{\epsffile{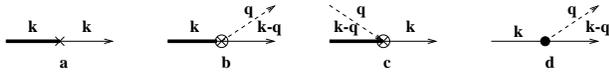}}
\caption{The vertices: a) - single hole creation, b) - hole + spin-wave
creation, c) - hole creation with spin wave annihilation, g) - usual 
hole-spin-wave vertex.
The thick line correspond to $G_c$, and the thin solid line corresponds to
$G_d$. The dashed line is the spin-wave.}
\label{fig:Fig3}
\end{figure}

At finite temperature there is another possibility: the creation of a hole
with the absorption of a spin wave from the thermal bath.
We denote this amplitude by $c_{\bf k,q}$. It is shown in figure~\ref{fig:Fig3}c,
and for simplicity we also denote it as a circled cross.
The derivation of $c_{\bf k,q}$ is quite similar to (\ref{bk}) and the
result is
\begin{equation}
\label{ck}
c_{\bf k,q}=\langle \alpha^{\dag}_{\bf -q}
d_{{\bf k-q}\downarrow} c_{{\bf k}\downarrow} \rangle
\approx\sqrt{2\over{N}} u_{\bf q}.
\end{equation}
In eqs. (\ref{bk}) and (\ref{ck}) we have omitted the standard 
Bose statistics factors $\sqrt{1+n_{\bf q}}$ in (\ref{bk}) and 
$\sqrt{n_{\bf q}}$ in (\ref{ck}). These factors are taken
into account separately in the calculation of the diagrams.

\begin{figure}
      \epsfxsize=80mm
      \centerline{\epsffile{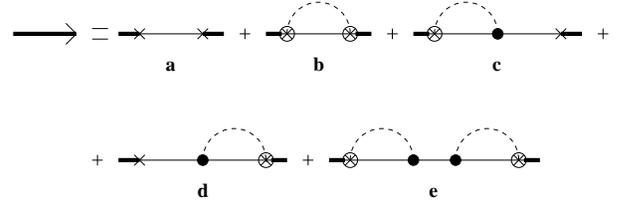}}
\caption{Dyson equation relating Green's functions $G_c$ (thick solid line)
and $G_d$ (thin solid line).}
\label{fig:Fig4}
\end{figure}

Now we can find the relation between the Green's functions $G_c$ (\ref{Gc}) 
and $G_d$ (\ref{Gd}). In leading $t$ approximation it is
given by the diagrams presented in figure~\ref{fig:Fig4} where the 
thin solid line represents the bare hole Green's function $G_{0d}$ 
(\ref{Gd0}). 
Each self-energy insertion should be understood as a combination of 
spin-wave emission and spin-wave absorption diagrams, similar to that in figure~\ref{fig:Fig1}.
Now let us dress these diagrams by higher orders in the hopping $t$.
As we already discussed above, there is no single loop correction to the 
``dot''.  We neglect double loop corrections to the ``dot''  as it has 
been done in SCBA. Therefore the only possibility is an introduction of
self energy corrections to $G_d$.
To take into account all these corrections we need just to replace
all bare hole Green's functions $G_{0d}$ by dressed
hole Green's functions $G_{d}$.
So, \ref{fig:Fig4} actually represents a Dyson equation relating $G_c$ (\ref{Gc}) 
and $G_d$ (\ref{Gd}). In analytical form it is
\begin{eqnarray}
\label{Dy2}
G_c(\epsilon,{\bf k})&=&a_{\bf k}^2 G_d(\epsilon,{\bf k})
+\Sigma_{1}(\epsilon,{\bf k}) \nonumber\\
&+&2 a_{\bf k} G_d(\epsilon,{\bf k}) \Sigma_{2}(\epsilon,{\bf k})
+G_d(\epsilon,{\bf k}) \Sigma_{2}^2(\epsilon,{\bf k}),
\end{eqnarray}
where
\begin{eqnarray}
\label{s1}
\Sigma_1(\epsilon,{\bf k})&=&\sum_{\bf q}
[(1+n_{\bf q})b^2_{\bf k-q,q} G_d(\epsilon -\omega_{\bf q},{\bf k-q}) \nonumber\\
&+&  n_{\bf q}c^2_{\bf k,q} G_d(\epsilon +\omega_{\bf q},{\bf k+q}) ],
\end{eqnarray}
and
\begin{eqnarray}
\label{s2}
\Sigma_2(\epsilon,{\bf k})&=&\sum_{\bf q}
[(1+n_{\bf q})b_{\bf k-q,q}g_{\bf k-q,q}
G_d(\epsilon-\omega_{\bf q},{\bf k-q}) \nonumber\\ &+& 
n_{\bf q}c_{\bf k,q}g_{\bf k,q}G_d(\epsilon +\omega_{\bf q},{\bf k+q}) ].
\end{eqnarray}

\subsection{Sum rules}

Let us check now the sum rules. All singularities of the retarded Green's functions
are in the lower half plane of complex $\epsilon$. Therefore if
we integrate eq.(\ref{Dy}) over $\epsilon$ from $-\infty$ to
$+\infty$, this integral can be replaced by the integral over an
infinite semi-circle in the upper $\epsilon$ half plane. For infinite
$\epsilon$, $G_d = G_{0d}$, and we get the well known sum rule
\begin{equation}
\label{sum}
-{1\over{\pi}} Im \int_{-\infty}^{\infty}G_d(\epsilon,{\bf k})
d \epsilon =1,
\end{equation}
which agrees with eq.(\ref{norm}).
If we integrate eq.(\ref{Dy2}) in the same limits, the terms
which contain more than one Green's function give no contribution,
because the integral can be transferred into the upper complex
$\epsilon$ half plane, and we find
\begin{eqnarray}
\label{sum1}
&&-{1\over{\pi}} Im \int_{-\infty}^{\infty}G_c(\epsilon,{\bf k})
d \epsilon =\left(
-{1\over{\pi}} Im \int G_d(\epsilon,{\bf k})
d \epsilon \right) \nonumber\\
&&\left(a_{\bf k}^2+ \sum_{\bf q}\left[
(1+n_{\bf q})b_{\bf k-q,q}^2 + n_{\bf q} c_{\bf k,q}^2\right]
\right)\nonumber\\
&&=0.5+{2\over{N}}\sum_{\bf q}\left[(1+n_{\bf q})v_{\bf q}^2
+ n_{\bf q}u_{\bf q}^2\right] \nonumber\\
&&= {2\over N}\sum_{\bf q}
{{A}\over{\omega_{\nu \bf q}}}(1+2n_{\bf q}) = 1,
\end{eqnarray}
where we have used eqs. (\ref{sw}) and (\ref{nu}).
Thus equation (\ref{Dy2}) reproduces the correct normalization:
$\langle 0|c^{\dag}_{{\bf k}\uparrow} c_{{\bf k}\uparrow}|0\rangle=1$. 
This also proves that the vertices (\ref{bk}) and (\ref{ck}) are correct. 

The vertices $b_{\bf k,q}$ (\ref{bk}) and $c_{\bf k,q}$ (\ref{ck}) are 
invariant under translation by the inverse vector of magnetic sub-lattice 
${\bf Q}=(\pm \pi,\pm \pi)$:
$b_{\bf k+Q,q}=b_{\bf k,q}$. At the same time the vertex $g_{\bf k,q}$
(\ref{g}) changes  sign with this translation:
$g_{\bf k+Q,q}=-g_{\bf k,q}$. Therefore the self energy 
$\Sigma_2(\epsilon,{\bf k})$ changes sign at ${\bf k} \to {\bf k+Q}$ and
\begin{equation}
\label{ne}
G_c(\epsilon,{\bf k+Q}) \ne G_c(\epsilon,{\bf k}).
\end{equation}
The imaginary part of $G_c(\epsilon,{\bf k})$ gives directly the spectra
measured in ARPES experiments. This Green's function can be calculated
using the Dyson equation (\ref{Dy2}) as soon as we have found $G_d$ in 
SCBA (\ref{Dy}).
\subsection{Results for G$_c$}

Numerical evaluation of $G_c$ at finite temperature, however, is more 
complicated than at T=0. 
The problem lies in the infrared divergence of the integrand of 
$\Sigma_1(\epsilon,{\bf k})$ at small ${\bf q}$.
To clarify this we compare the small q behavior of the integrands of the 
self-energies $\Sigma$, $\Sigma_1$, and $\Sigma_2$.
Small q means that $1/\xi_M \ll q \ll T$, where $\xi_M \propto \exp(1.1/T)$ is 
the magnetic
correlation length. In this region the spin-wave mean occupation number is
$n_{\bf q} \sim T/q$, and the vertices are $g_{\bf k,q} \sim \sqrt{q}$,
$b_{\bf k,q} \approx c_{\bf k,q} \sim 1/\sqrt{q}$.
The self energy $\Sigma$ has an integrand
$\propto n_{\bf q}g_{\bf k,q}^2 d^2q \sim T q dq$. It is convergent at small q
and therefore numerical calculation of $\Sigma$ is straightforward and the 
finite temperature generalization of SCBA is as simple as zero temperature SCBA.
For the integrand of the self-energy $\Sigma_2$ on finds 
$\propto n_{\bf q}b_{\bf k,q} g_{\bf k,q} d^2q \sim T dq$, which is also 
convergent at small q.
The situation is different in self-energy $\Sigma_1$: 
it is logarithmically divergent at small q:
$\propto n_{\bf q}b_{\bf k,q}^2 d^2q \sim T dq/q$. 
There is  no real divergence, however, because the integral is convergent 
at $q \sim 1/\xi_M$,
but to calculate this integral numerically by ``brute force'' one needs 
a grid with $\Delta q \ll 1/\xi_M$. One then needs for example, 
when T=0.25, a lattice of at least $200 \times 200$, and for lower 
temperatures an even a bigger lattice. 

\begin{figure}
      \epsfxsize=80mm
      \centerline{\epsffile{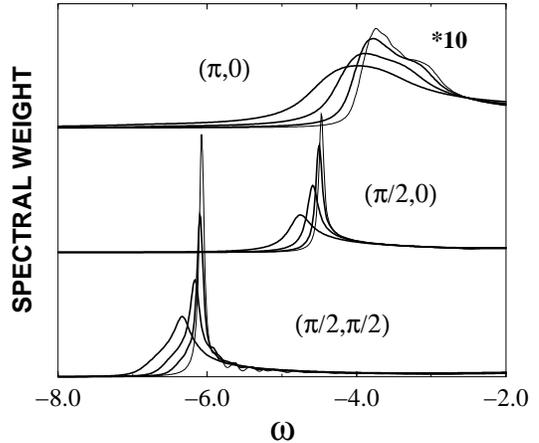}}
\caption{Plots of $-{1\over{\pi}} \ Im \ G_c(\omega, {\bf k})$
for different temperatures and  {\bf k}-points. The thin line is the result for $T=0.01$,
the thick lines for $T=0.2,0.3$ and $0.4$. The spectral weight at the $(\pi,0)$ point
is multiplied by a factor 10.}
\label{fig:g_c}
\end{figure}

We can avoid this problem by renormalizing $\Sigma_1$, so that we
can work with a reasonable grid-size. 
Let us rewrite eq. (\ref{s1}) in the form
\begin{equation}
\label{s1r}
\Sigma_1(\epsilon,{\bf k})=\Sigma_R(\epsilon,{\bf k})+
\Sigma_{RR}(\epsilon,{\bf k}),
\end{equation}
where
\begin{eqnarray}
\label{sr}
\Sigma_R(\epsilon,{\bf k})&=&\sum_{\bf q}
(1+n_{\bf q})b^2_{\bf k-q,q}\nonumber \\
&&\left[ G_d(\epsilon -\omega_{\bf q},{\bf k-q})-G_d(\epsilon,{\bf k})\right]\\
&+& \sum_{\bf q}n_{\bf q}c^2_{\bf k,q} \left[ G_d(\epsilon
+\omega_{\bf q},{\bf k+q})- G_d(\epsilon,{\bf k})\right],\nonumber
\end{eqnarray}
and
\begin{equation}
\label{srr}
\Sigma_{RR}(\epsilon,{\bf k})= G_d(\epsilon,{\bf k})\sum_{\bf
q}\left[(1+n_{\bf q})b^2_{\bf k-q,q} + n_{\bf q}c^2_{\bf k,q} \right].
\end{equation}
Numerical calculation of $\Sigma_{R}$ does not cause any trouble
because it is well convergent at small q. On the other hand
$\Sigma_{RR}$ can be easily calculated analytically using the modified
spin-wave theory equations (\ref{sw}) and (\ref{nu}):
$\Sigma_{RR}(\epsilon,{\bf k})= {1\over{2}} G_d(\epsilon,{\bf k})$.
Using this procedure the calculation can be done at arbitrary small 
temperature. The results are practically independent of the grid as soon
as it is larger  than $20 \times 20$.
The plots of $-{1\over{\pi}} \ Im \ G_c(\epsilon, {\bf k})$
as a functions of $\epsilon$ for ${\bf k}=(\pi/2, \pi/2)$,
${\bf k}=(\pi/2,0)$, and ${\bf k}=(\pi,0)$
are presented in figure~\ref{fig:g_c}.

The half widths of quasiparticle peaks of $G_c$ are 
slightly larger than the half widths of quasiparticle peaks of $G_d$.
The spectra at T=0.01 quite well agrees
with zero temperature calculation~\cite{8G1}. We stress that the agreement
is not trivial. At T=0 long range anti-ferromagnetic order
is assumed, and  in the present work we used a quite
different approach based on a state without long range order.
The agreement indicates that these two approaches are consistent.
For non-zero temperatures the trend in the spectra for $G_c$ is the same
as for $G_d$; the temperature effect is to shift the quasi-particle peaks
uniformly to lower energies and to broaden them.
The spectra presented in figure~\ref{fig:g_c} should be directly compared with ARPES 
experimental data~\cite{8Wells,8Pothuizen}. They reasonably reproduce
positions and residues of experimental peaks, but fail to reproduce
widths of the peaks.

\section{Conclusions}
We considered the two-dimensional $t-t^{\prime}-t^{\prime \prime}-J$ model 
at finite temperature,
and developed a technique to deal with the state without long range
anti-ferromagnetic order. There is hope to extend this technique 
to the spin liquid state of the doped system.
We generalized the self-consistent Born
approximation to the case of nonzero temperature and derived the
Dyson equation which relates the single hole Green's function with 
fixed spin to the single hole Green's function with fixed pseudo-spin.
This equation is sensitive to very large distances of order of magnetic
correlation length and therefore not convenient for
computations. To overcome this problem we developed  a renormalization 
procedure which allows one to exclude large distances and to work with
a relatively small lattice: the results are independent of the lattice
size as soon as it is larger than $20 \times 20$.
The effect of a finite temperature is a broadening and a shift of 
the quasi-particle peaks to lower energy, independent of the momentum. 
This is attributed to the frustrated
magnetic order at finite temperature. 
The calculated ARPES spectra demonstrate that temperature broadening is not
enough to explain widths of the experimental spectra~\cite{8Wells,8Pothuizen}.
This strengthens the argument that other degrees of freedom contribute to
the peak width~\cite{8G1}.

\section{Acknowledgments}
We are very grateful to M. Kuchiev and G.A. Sawatzky for stimulating discussions.
This work was financially supported by the Nederlandse Stichting voor Fundamenteel 
Onderzoek der Materie (FOM) and the Stichting Scheikundig Onderzoek Nederland (SON),
both financially supported by the Nederlandse Organisatie voor Wetenschappelijk
Onderzoek (NWO).



\begin{thebibliography}{99}
 
\bibitem[a]{G} Materials Science Center, University of Groningen,
Nijenborgh 4, 9747 AG Groningen, The Netherlands.
\bibitem[b]{Budker} Also at the Budker Institute of Nuclear Physics,
  630090 Novosibirsk, Russia

\bibitem{8Wells} B. O. Wells {\it et al}, Phys. Rev. Lett. {\bf 74}, 964 (1995).
\bibitem{8Pothuizen} J. J. Pothuizen {\it et al}, Phys. Rev. Lett. {\bf 78},
717 (1997).
\bibitem{8Naz} A. Nazarenko, K. J. E. Vos, S. Haas, E. Dagotto, and
R. J. Gooding, Phys. Rev. B {\bf 51}, 8676 (1995).
\bibitem{8Bala} J. Bala, A. M. Oles, and J. Zaanen, Phys. Rev. B
{\bf 52} 4597 (1995).
\bibitem{8And} O. K. Andersen {\it et al}, J. Phys. Chem. Solids {\bf 56}
1573 (1995).
\bibitem{8G1} O. P. Sushkov, G. Sawatzky, R. Eder, and H. Eskes,
submitted to Phys. Rev. B. 
\bibitem{8Taka} M. Takahashi Phys. Rev. B {\bf 40}, 2494 (1989).
\bibitem{8Tok} Y. Tokura {\it et al}, Phys. Rev. B {\bf 41}, 11657 (1990).
\bibitem{8Grev} M. Greven {\it et al}, Phys. Rev. Lett. {\bf 72} 1096
(1994).
\bibitem{8Schmitt} S. Schmitt-Rink, C. M. Varma, and 
A. E. Ruckenstein, Phys. Rev. Lett. {\bf 60}, 2793 (1988).
\bibitem{8Kane} C. L. Kane, P. A. Lee, and N. Read, Phys. Rev. B {\bf 39},
6880 (1989).
\bibitem{8Mart} G. Martinez and P. Horsch, Phys. Rev. B {\bf 44}, 317 (1991).
\bibitem{8Liu} Z. Liu and E. Manousakis, Phys. Rev. B {\bf 45}, 2425 (1992).
\bibitem{8Susf} O. P. Sushkov, Phys. Rev. B {\bf 49}, 1250 (1994).
\bibitem{8Manousakis} E. Manousakis, Rev. Mod. Phys. {\bf 63}, 1 (1991).
\bibitem{8DM}F. J. Dyson, Phys. Rev. {\bf 102}, 1217,1230 (1956);
S. V. Maleev, Zh. Eksp. Teor. Fiz. {\bf 30}, 1010 (1957) [Sov. Phys.
JETP {\bf 6}, 776 (1958).
\bibitem{8Ed} R. Eder, Y. Ohta, and  G. Sawatzky, Phys. Rev. B {\bf 55}, R3414 
(1997). 

\end{thebibliography}
\end{document}